\pdfoutput=1
\documentclass[pdflatex,sn-nature]{sn-jnl}% Style for submissions to Nature Portfolio journals
\usepackage{graphicx}%
\usepackage{multirow}%
\usepackage{amsmath,amssymb,amsfonts}%
\usepackage{amsthm}%
\usepackage{mathrsfs}%
\usepackage[title]{appendix}%
\usepackage{xcolor}%
\usepackage{textcomp}%
\usepackage{manyfoot}%
\usepackage{booktabs}%
\usepackage{algorithm}%
\usepackage{algorithmicx}%
\usepackage{algpseudocode}%
\usepackage{listings}%
\usepackage{float}
\newcommand{\blyr}{$\beta$\,Lyr}

\theoremstyle{thmstyleone}%
\theoremstyle{thmstyletwo}%

\theoremstyle{thmstylethree}%

\begin{document}

\title[Mauve Observations of $\beta$ Lyrae]{Mauve Observations of the Eclipsing Binary System $\beta$ Lyrae}


\author*[1]{\fnm{Sharafina} \sur{Razin}}\email{sharafina@bssl.space}

\author[1]{\fnm{Imogen} \sur{Garlick}}
\author[1,2,3]{\fnm{Fabio} \sur{Favata}}
\author[1,4]{\fnm{Jonathan} \sur{Tennyson}}
\author[1,5]{\fnm{Arianna} \sur{Saba}}
\author[1,5]{\fnm{Giovanna} \sur{Tinetti}}
\author[1]{\fnm{Richard} \sur{Archer}}
\author[1]{\fnm{Lawrence} \sur{Bradley}}
\author[1]{\fnm{Sarah} \sur{Harvey}}
\author[1]{\fnm{Gabriella} \sur{Hodosán}}
\author[1]{\fnm{Harry} \sur{Johnson}}
\author[1]{\fnm{James} \sur{McLaren}}
\author[1]{\fnm{Giorgia} \sur{Parente}}
\author[1]{\fnm{Yoga Barrathwaj} \sur{Raman Mohan}}
\author[1,4]{\fnm{Giorgio} \sur{Savini}}
\author[1]{\fnm{David} \sur{de Sousa Seaman}}
\author[1]{\fnm{Ian} \sur{Stotesbury}}
\author[1]{\fnm{Marcell} \sur{Tessenyi}}
\author[1]{\fnm{Benjamin J.} \sur{Wilcock}}
\author[1]{\fnm{Philip} \sur{Windred}}

%%%%%%%%%%%%%%
% Collaboration
\author{\fnm{Mauve Science Collaboration - Year 1:}}
\author[6]{\fnm{Marcel A.} \sur{Agüeros}}
% \author[7]{\fnm{Don} \sur{Dixon}}
\author[8,9]{\fnm{Chuanfei} \sur{Dong}}
\author[7]{\fnm{Girish M.} \sur{Duvvuri}}
\author[10]{\fnm{Patrick F.} \sur{Flanagan}}
\author[11]{\fnm{Christopher M.} \sur{Johns-Krull}}
\author[12]{\fnm{Theo \sur{Khouri}}}
\author[8]{\fnm{Hongpeng} \sur{Lu}}
\author[13]{\fnm{Hiroyuki} \sur{Maehara}}
\author[14,15,16,17]{\fnm{Kosuke} \sur{Namekata}}
\author[6]{\fnm{Alejandro} \sur{Núñez}}
\author[18]{\fnm{Elena} \sur{Pancino}}
\author[18]{\fnm{Sharmila} \sur{Rani}}
\author[19]{\fnm{Anusha} \sur{Ravikumar}}
\author[21,22]{\fnm{Blanka}\sur{Schmercz}}
\author[21]{\fnm{Bálint} \sur{Seli}}
\author[19, 20]{\fnm{T. A. A.} \sur{Sigut}}
\author[7]{\fnm{Keivan G.} \sur{Stassun}}
\author[10]{\fnm{Jamie J.} \sur{Stewart}}
\author[21]{\fnm{Krisztián} \sur{Vida}}
\author[12]{\fnm{Wouter} \sur{Vlemmings}}
\author[10]{\fnm{Emma T.} \sur{Whelan}}

% Affiliations
\affil*[1]{\orgname{Blue Skies Space Ltd.}, \orgaddress{\street{69 Wilson Street}, \city{London}, \postcode{EC2A 2BB}, \country{UK}}}

\affil[2]{\orgname{INAF – Osservatorio Astronomico di Palermo}, \orgaddress{\street{Piazza del Parlamento 1}, \city{Palermo}, \postcode{90134}, \country{Italy}}}

\affil[3]{\orgdiv{Department of Physics}, \orgname{Imperial College London}, \orgaddress{\street{Exhibition Road}, \city{London} \postcode{SW7 2AZ}, \country{UK}}}

\affil[4]{\orgdiv{Department of Physics and Astronomy}, \orgname{University College London}, \orgaddress{\street{Gower Street}, \city{London}, \postcode{WC1E 6BT}, \country{UK}}}

\affil[5]{\orgdiv{Department of Physics}, \orgname{King’s College London}, \orgaddress{\street{Strand}, \city{London}, \postcode{WC2R 2LS}, \country{UK}}}

\affil[6]{\orgdiv{Department of Astronomy}, \orgname{Columbia University}, \orgaddress{\street{550 West 120th Street}, \city{New York}, \postcode{NY 10027}, \country{USA}}}

\affil[7]{\orgdiv{Department of Physics and Astronomy}, \orgname{Vanderbilt University}, \orgaddress{\city{Nashville}, \postcode{TN 37235}, \country{USA}}}

\affil[8]{\orgdiv{Department of Astronomy}, \orgname{Boston University}, \orgaddress{\city{Boston}, \postcode{MA 02215}, \country{USA}}}

\affil[9]{\orgdiv{School of Natural Sciences}, \orgname{Institute for Advanced Study}, \orgaddress{\city{Princeton}, \postcode{NJ 08540}, \country{USA}}}

\affil[10]{\orgdiv{Department of Physics}, \orgname{Maynooth University}, \orgaddress{\city{Maynooth}, \postcode{Co Kildare}, \country{Ireland}}}

\affil[11]{\orgdiv{Department of Physics and Astronomy}, \orgname{Rice University}, \orgaddress{\street{6100 Main Street}, \city{Houston}, \postcode{TX 77005}, \country{USA}}}

\affil[12]{\orgdiv{Department of Physics and Astronomy}, \orgname{Chalmers University of Technology}, \postcode{412 96}, \city{Gothenburg}, \country{Sweden}}

\affil[13]{\orgname{Subaru Telescope Okayama Branch Office, National Astronomical Observatory of Japan}}

\affil[14]{\orgname{The Hakubi Center for Advanced Research, Kyoto University}, \orgaddress{\city{Kyoto}, \postcode{606-8501}, \country{Japan}}}

\affil[15]{\orgdiv{Department of Physics}, \orgname{Kyoto University}, \orgaddress{\city{Kyoto}, \postcode{606-8502}, \country{Japan}}}

\affil[16]{\orgname{Heliophysics Science Division, NASA Goddard Space Flight Center}, \orgaddress{\city{Greenbelt}, \postcode{MD 20771}, \country{USA}}}

\affil[17]{\orgname{The Catholic University of America}, \orgaddress{\city{Washington}, \postcode{DC 20064}, \country{USA}}}

\affil[18]{\orgname{INAF - Osservatorio Astrofisico di Arcetri}, \orgaddress{\city{Firenze}, \postcode{50125}, \country{Italy}}}

\affil[19]{\orgdiv{Department of Physics \& Astronomy}, \orgname{The University of Western Ontario}, \orgaddress{\city{London}, \postcode{ON N6A 3K7}, \country{Canada}}}

\affil[20]{\orgdiv{Institute for Earth and Space Exploration (IESX)}, \orgname{The University of Western Ontario}, \orgaddress{7134 Perth Drive, London, \postcode{ON N6A 5B7}, \country{Canada}}}

\affil[21]{\orgdiv{HUN-REN Research Centre for Astronomy and Earth Sciences}, \orgname{Konkoly Observatory}, \orgaddress{MTA Centre of Excellence, Konkoly Thege Miklós út 15-17., \postcode{H-1121} \city{Budapest}, \country{Hungary}}}

\affil[22]{\orgdiv{Department of Astronomy}, \orgname{Eötvös University}, \orgaddress{Pf. 32, \postcode{1518}, \city{Budapest}, \country{Hungary}}}

\abstract{We present a broadband, synoptic observing campaign of the complex eclipsing binary system \blyr, obtained with the first commercial astronomical satellite, Mauve, developed and operated by Blue Skies Space Ltd. Mauve's capabilities have resulted in the first multiphase monitoring of \blyr, covering UV and optical photometry, as well as low-resolution spectrophotometry, including the phase-folded mapping of the intensity of a number of emission lines. The different components in its spectrum vary in very different ways, indicative of a system with a number of physical components. Our data show that while most of the components show repeatable cycle-to-cycle variability, the thus far hardly observed UV emission (2000-3000\,\AA) varies significantly from one cycle to the other, pointing to the need for simultaneous observations to understand and model the system. These observations reveal that the ultraviolet-emitting structures in this benchmark interacting binary evolve on timescales of individual orbits, challenging the assumption of a stationary accretion geometry commonly adopted in models of interacting binary stars. The quality of the data set and the fresh view that they bring to a well studied, but still poorly understood object demonstrate the scientific value of small telescopes with focused scientific capabilities, as well as the value of a new generation of commercial astronomical instrumentation.}


\maketitle

Stars in short period binaries can follow a distinctly different evolutionary path than single stars or even stars in wide binary systems \citep{1971ARA&A...9..183P}. If the two stars orbit sufficiently close to each other when the evolution of the more massive star increases its size due to core hydrogen exhaustion, filling its Roche lobe, the overflow through the Lagrangian point will result in mass transfer toward the less massive component. At this point, evolution of both members of the binary system will be heavily affected, with the more massive component losing mass and the less massive component accreting it, possibly through an accretion disc \citep{1975ApJ...198..383L}. This can lead to the so-called Algol paradox \citep{1955ApJ...121...71C}, i.e., binary systems in which the less massive star appears to be more evolved than more massive star in the system, contrary to standard stellar evolution, in which more massive stars evolve more rapidly. The bright eclipsing binary Algol is the best known such system \citep{2012ApJ...752...20B}. Eclipsing binaries are fundamental benchmarks for stellar astrophysics: their orbits and eclipses yield direct, distance-independent measurements of stellar masses and radii, against which models of stellar evolution can be tested \citep{2004ApJS..151..357S, 2006Natur.440..311S}.

$\beta$ Lyrae is a  textbook example of a binary system caught in the short, rapid mass-transfer phase of close-binary evolution, with a $\sim$3 $M_{\odot}$ donor feeding a $\sim$13 $M_{\odot}$ gainer. The favourable viewing geometry makes it an eclipsing system, enabling the study of the system’s dynamics. While $\beta$ Lyrae has been studied extensively, and for a long time it has been considered as a paradigmatic contact binary, it is a complex system  with several yet unexplained photometric and spectral peculiarities. ``Burnham's celestial handbook'' in 1978 \cite{1978bchv.book.....B} stated that ``$\beta$ Lyrae is one of the most frustratingly interesting stars in the sky to the astrophysicist, and has probably been the subject of more studies and investigations than any other star in the heavens with the single exception of our own Sun.'' Remarkably, for a system of such paradigmatic nature, very few observations in the ultraviolet (UV) exist, the most recent \cite{1994ApJ...421..787K} dating back to the International Ultraviolet Explorer (IUE) which flew between 1978 and 1996. No campaign simultaneously covering the UV and visible band has been carried out prior to the observations herewith reported\footnote{a handful of UV observations, apparently unpublished, are present in the HST archive.}. 

$\beta$ Lyrae's  spectrum shows strong emission lines in the Balmer series and from He\,I, showing significant non-photospheric emission which has led to its being also considered a member of the Be star group. The ongoing mass transfer is also inferred from the secular increase of the orbital period, 19 seconds per year.

The current "standard model" for $\beta$ Lyrae has several components, i.e., the visible primary star, an optically thick disc entirely hiding the secondary star, a large bipolar jet protruding from the disc (not centered on the hidden secondary) and a large circumstellar envelope (e.g., \citep{2007A&A...463..233A}, \citep{2002AN....323...87H}). Each component contributes to the system's emission, so that its spectral energy distribution (SED) will be very different from the simple superposition of two photospheres. Thermal (photospheric) emission will come from the visible primary and from the disc, while the emission lines will likely come from the optically thin jets and from the extended envelope. The disc is also thought to have a possible hot spot due to the accretion shock, which can produce a UV excess.

The system was observed in the UV by both the Orbiting Astronomical Observatory 2 (OAO-2, 1968-73) and the IUE satellites, as discussed by Kondo et al. \citep{1994ApJ...421..787K}, who show that the eclipses disappear in the far UV, blueward of the Ly$\alpha$ line, while red-ward of about 2000\,\AA\ the eclipses are similar to the ones observed in the optical. Their interpretation sees the far UV emission coming from a putative large circumstellar envelope, possibly becoming optically thick at the shorter wavelengths (and therefore with its luminosity being unaffected by the stars orbiting inside it). Kondo et al. state that they have "no specific model" for this putative envelope. 

Thus far, no single model (even with the several components listed above) appears capable of explaining all the features of the system. Some authors have even questioned whether the invisible secondary hidden inside the accretion disc is a compact, degenerate object rather than a normal star. While most current work assumes that the hidden accretor is a normal star, the only X-ray observation \cite{2008A&A...477L..37I}, taken by the Suzaku Observatory in 2006, show strong soft X-ray emission remaining constant in phase and with no evidence of eclipses (similarly to the UV flux blue-ward of Ly$\alpha$). The X-ray spectrum is soft and compatible with a thermal source at $T \simeq 0.6$\,keV, which Ak et al. \cite{2008A&A...477L..37I}  attribute to the (scattered) wind from the hidden accretor, but intriguingly  also claim evidence for a hard X-ray tail in excess of 10\,keV, which would indeed be compatible with a compact object.

The last decade has witnessed a renewed interest in the star: \citep{2018A&A...618A.112M} (Mea2018) and \citep{2013MNRAS.432..799M} (MD2013)  carried out extensive optical observations on which they based a detailed model of the system, in which the star accreting mass from the donor is hidden within a large, optically thick accretion disc, which features a hot spot. The two models feature similar stellar masses and disc sizes, with somewhat different inclination, although in both cases the disc is seen almost edge-on, with MD2013 modelling it as a thick slab of material (6.5\,$R_\odot$) hiding the gainer star and contributing $\sim$22\% of the V-band light at quadrature. The disc does not appear to be in vertical hydrostatic equilibrium, indicating ongoing mass transfer, and a hot spot is required to fit the models. Both studies list a full set of orbital and physical parameters of $\beta$\,Lyrae. 

$\beta$\,Lyr's brightness makes it an ideal target for small telescopes, and for example  a long photometric observing campaign in visible light has been carried out with cubesats \citep{2018AJ....156...12R}. Mauve is a small satellite developed by Blue Skies Space Ltd. with observing programmes set by scientists from participating institutions \cite{10.1093/astrogeo/ataf048}. Mauve carries a 13-cm UV-visible (2000-7000 \AA) spectrophotometric space telescope, and is a perfect instrument for the study of $\beta$ Lyrae, given its simultaneous coverage of a broad wavelength range spanning from the UV to the red and its capacity to resolve strong emission lines. We conducted a Mauve observing campaign of $\beta$ Lyrae running from the 5th of August to the 4th of the October 2026, during which 93 observations with good phase coverage were obtained. Attempts at modelling the $\beta$\,Lyr system in the past had to rely on data sets which were taken at widely different times, e.g., reconciling the decade-old UV light curves with a set of later photometric light curves in the visible and yet another set of emission lines observations. Mauve's data provide the first opportunity to base a model of the system on data acquired simultaneously. An extensive description of the mission is available \cite{10.1093/astrogeo/ataf048, 2025RASTI...4af045S}.

\begin{figure}[!ht]
    \centering
    \includegraphics[width=0.9\textwidth]{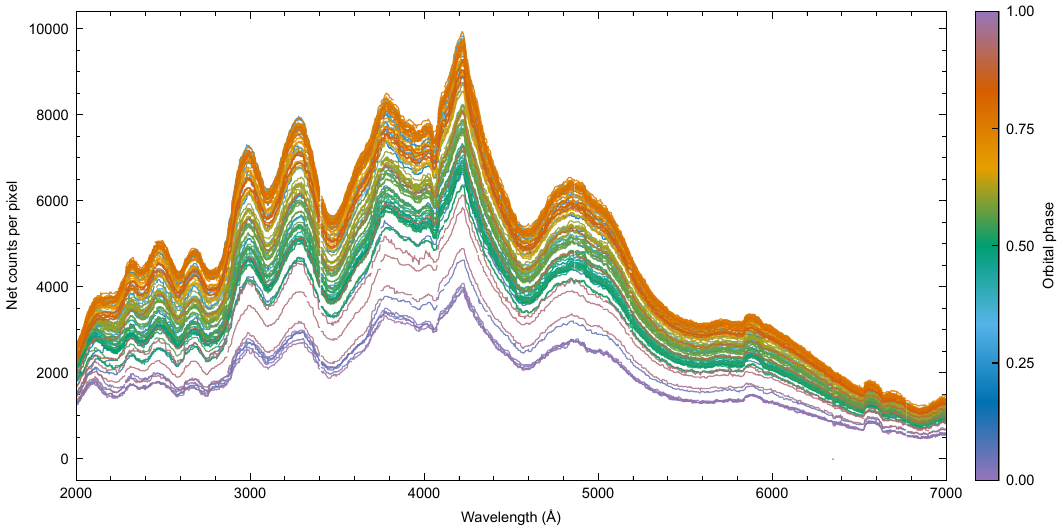}
    \caption{Mauve spectra of $\beta$~Lyr from 93 observations covering the full orbital period, shown as net counts per pixel against wavelength. Each spectrum is coloured by the orbital phase at which it was taken (colour bar). The orbital phase is the fraction of the 12.94-d orbital period elapsed since primary minimum ($\phi=0$, when the donor is eclipsed by the disc), computed from the quadratic ephemeris of \cite{2007A&A...463..233A}. Spectra are smoothed with a 7-pixel median filter; dotted lines mark the main emission lines.}
    \label{fig:netcounts}
\end{figure}

The 93 background-subtracted spectra are shown in Fig.~\ref{fig:netcounts}, where a strong phase-modulation of more than a factor of 2 is evident.

The spectra were analysed through both a synthetic photometry approach, deriving magnitudes in a number of bands, and by looking for, and determining the equivalent width, of key emission lines (H$\alpha$, H$\beta$, H$\gamma$, He\,I~5876\,\AA\,\ and He\,I~6678\,\AA). 
This analysis was performed by comparison with standard stars, Vega for the photometry and the standard B7 star 29\,Psc for the spectroscopic analysis. 29\,Psc (HR\,9087) is a B7II giant star and a close spectral match to $\beta$\,Lyrae's donor (the component that dominates the spectrum). 

%\section*{Results} \label{sec:results}

\begin{figure}[!ht]
    \centering
    \includegraphics[width=0.9\textwidth]{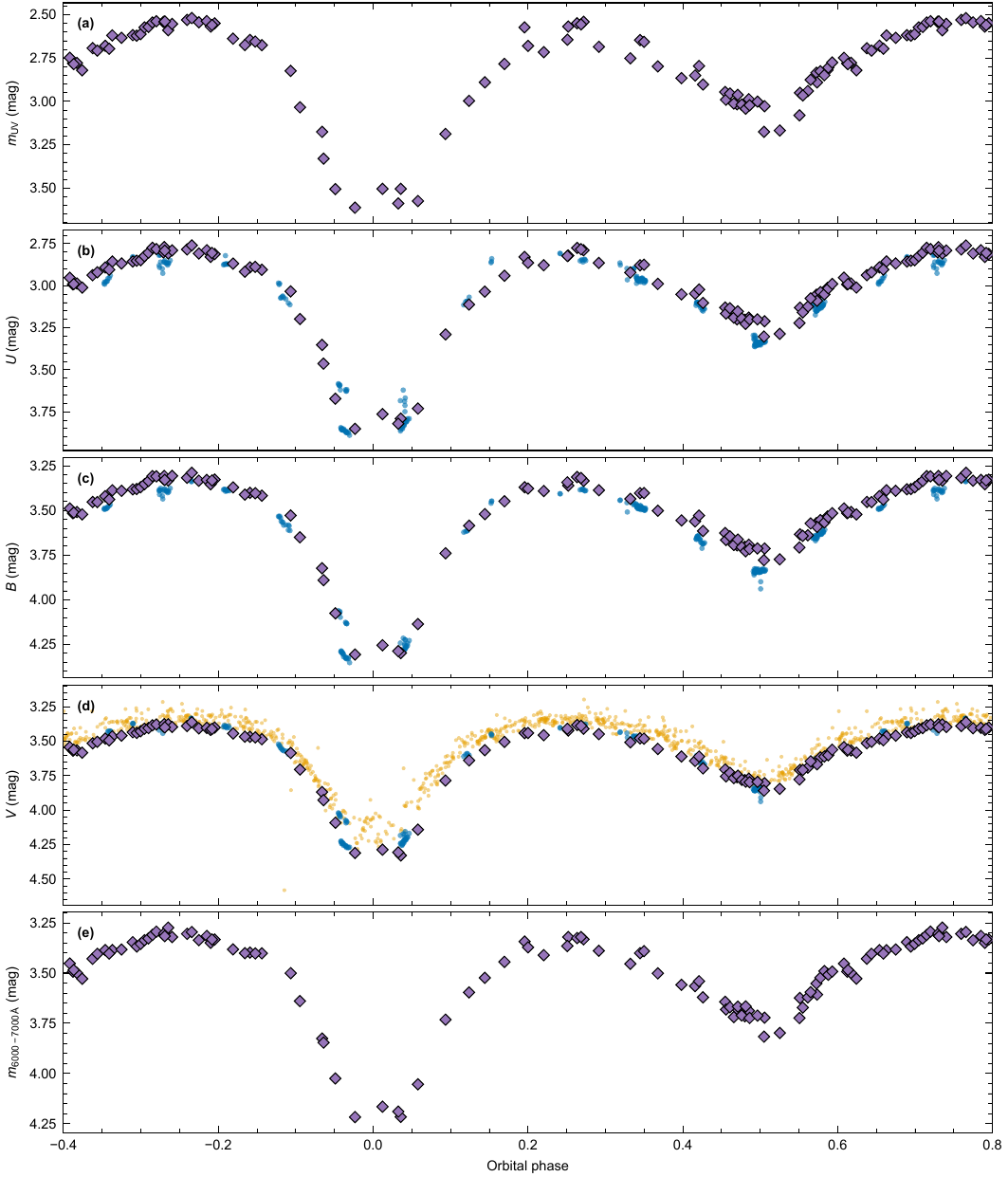}
    \caption{Synthetic photometry light curves of $\beta$~Lyrae from Mauve spectra (purple diamonds) in a 2000--3000\,\AA\ ($m_{UV}$) top-hat band (a), the Johnson $U$, $B$ and $V$ bands (b--d) and a 6000--7000\,\AA\ top-hat band (e), compared with photometry from Ak et al.\  \cite{2007A&A...463..233A} (blue dots) and Van Hamme et al.\ \cite{1995AJ....110.1350V} (yellow dots). Mauve error bars are statistical only; integrating the spectra over each band makes them mostly smaller than the symbols.}
    \label{fig:phot}
\end{figure}

We derived Johnson UBV magnitude for each of the spectra, together with magnitude in the 2000--3000\,\AA\ ($m_{\rm UV}$) and 6000--7000\,\AA\ band. Phase-folded light curves are shown in Fig.\,\ref{fig:phot}, together with  literature data \cite{1995AJ....110.1350V, 2007A&A...463..233A} for the UBV bands. The Mauve observations show a high degree of repeatability and of consistency with previous data in the standard UBV colours.

The primary eclipse shows an irregular shape, more visible in the U and $m_{\rm UV}$ bands. The spread of the photometric measurements of $\beta$\,Lyr during the primary minimum in the literature is large (e.g., the data of Van Hamme et al. \cite{1995AJ....110.1350V} shown in Fig,\,\ref{fig:phot}). Given that the accretion disc is a dynamic structure, its variability is not surprising, so that the profile of the primary eclipse has limited reproducibility across orbital cycles. Mea2018 \cite{2018A&A...618A.112M} find that a hot spot in the disc is necessary to provide a good fit to the V-band lightcurve, as previously discussed \citep{2013MNRAS.432..799M,Lomax_2012}. Notably, Mea2018's high-quality V-band series of photometric data, covering many orbital cycles, has a large spread in the primary minimum. The model of Mea2015 \citep{2013MNRAS.432..799M} also makes use of a hot spot to model the clear asymmetry in the light curve around the secondary eclipse, that again implies a structure in the disc.

\begin{figure}[!ht]
    \centering
    \includegraphics[width=0.9\textwidth]{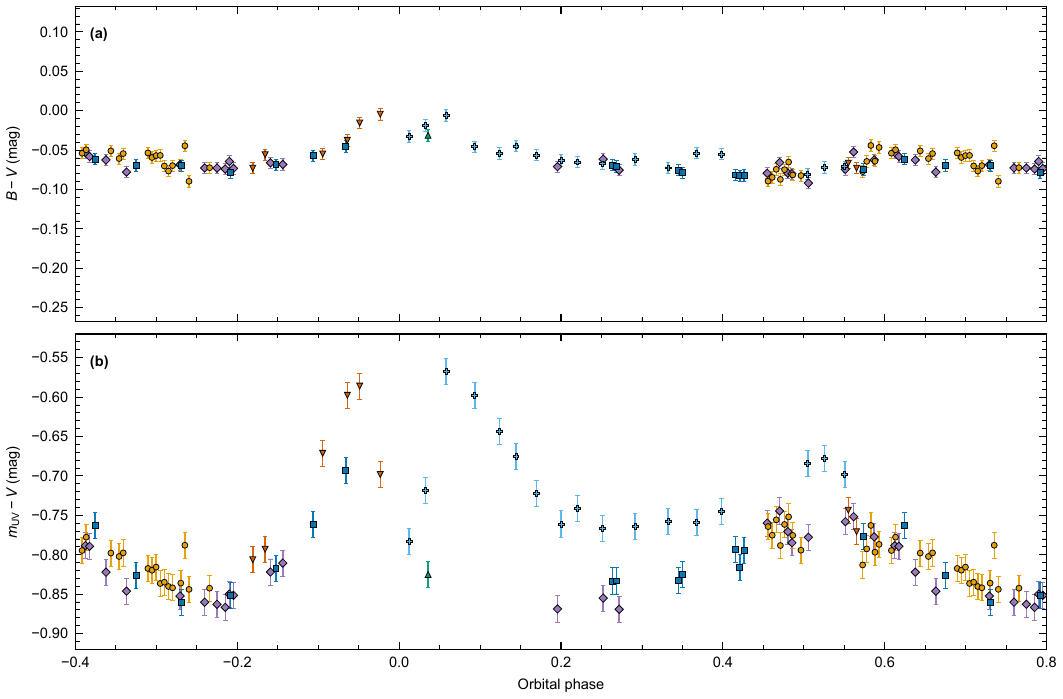}
    \caption{Colour curves from the Mauve synthetic photometry: (a) $B-V$, (b) $m_\mathrm{UV}-V$, where $m_\mathrm{UV}$ is the 2000--3000\,\AA\ top-hat magnitude. Colour and marker show the orbital cycle: [purple diamonds: cycle 1 (25 Jul--7 Aug); orange circles: cycle 2 (7 Aug--20 Aug); blue squares: cycle 3 (20 Aug--1 Sep); green triangles: cycle 4 (1 Sep--14 Sep); orange inverted triangles: cycle 5 (14 Sep--27 Sep); light blue crosses: cycle 6 (27 Sep--10 Oct)]. Error bars combine the statistical errors of the two magnitudes in quadrature with the scatter between observations at the same phase. Both panels span 0.4\,mag.}
    \label{fig:color_plot}
\end{figure}

We have analysed the variation in phase of two colours, i.e., the standard $B-V$ and the ad hoc colour $m_{{\rm UV}}-V$ (Fig.\,\ref{fig:color_plot}). The $B-V$ colour shows small but well-defined variations in phase, with the system slightly bluer during primary eclipse (implying a somewhat hotter region in the accretion disc, visible at primary eclipse), the $m_{{\rm UV}}-V$ colour is strongly modulated, with a strong enhancement at primary eclipse, but more importantly, significant period-to-period variability, never seen before in \blyr. The most likely interpretation of this strong variability is that the structure responsible for the UV emission is highly time-variable, and may not always be present.

\begin{figure}[!ht]
    \centering
    \includegraphics[width=0.9\textwidth]{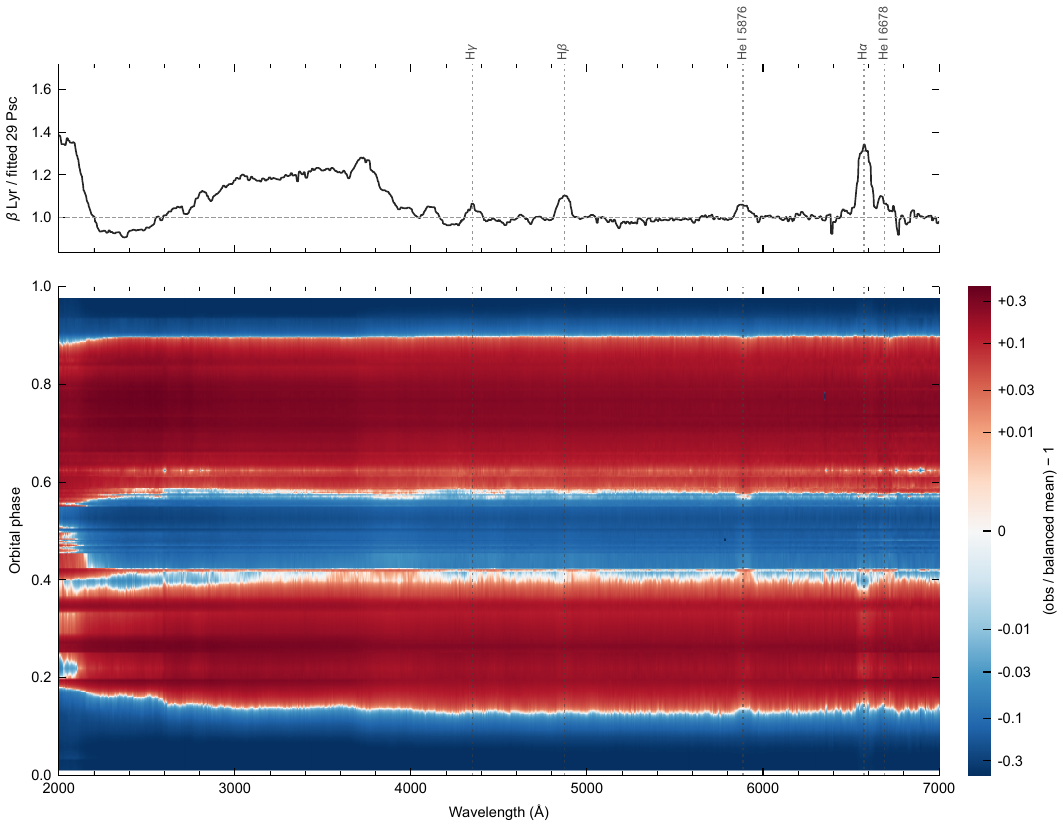}
    \caption{\textit{Top:} ratio of the phase-averaged $\beta$~Lyr spectrum (each phase bin weighted equally) to the fitted 29~Psc spectrum, with H$\gamma$, H$\beta$, He\,\textsc{i}~5876, H$\alpha$ and He\,\textsc{i}~6678 marked by dotted lines. The ratio shows a broad excess at $\sim$2800--4000~\AA\ and emission at the Balmer and He\,\textsc{i} lines, strongest at H$\alpha$. \textit{Bottom:} fractional deviation of each spectrum from the phase-averaged spectrum as a function of orbital phase and wavelength (symmetric logarithmic colour scale). The pattern follows the orbital light curve, with minima near phases 0.0 and 0.5} 
    \label{fig:29psc_fit_log}
\end{figure}

\begin{figure}[!ht]
    \centering
    \includegraphics[width=0.9\textwidth]{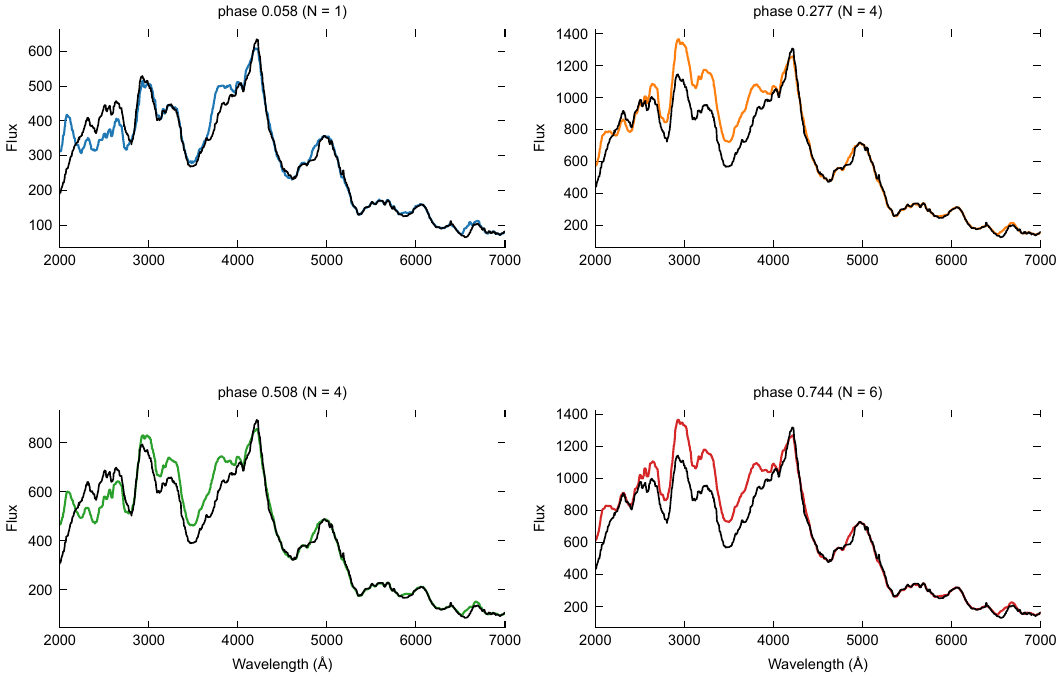}
    \caption{Mean Mauve spectra of $\beta$~Lyrae at orbital phases near primary eclipse (blue; 0.058), first quadrature (orange; 0.277), secondary eclipse (green; 0.508), and second quadrature (red; 0.744), each the mean of the $N$ spectra within $\pm0.02$ in phase, compared with the spectrum of 29~Psc (black; mean of eight observations) scaled to each $\beta$~Lyrae spectrum by a first-order polynomial in wavelength. The UV variations across the different phase can clearly be seen.}
    \label{fig:29psc_fit_phases} 
\end{figure}

We studied the spectral variability throughout the phase by comparing the spectrum with the one of 29\,Psc. We normalised the spectrum of $\beta$\,Lyr to the one of 29\,Psc in the region red-ward of 4000\,\AA\ and produced a ''difference spectrum'', both for the average spectrum and for each quadrature and conjunction, subtracting one from the other. Fig.\,\ref{fig:29psc_fit_log} shows the phase-folded variation of the Mauve spectra of $\beta$ Lyrae throughout the orbit, normalised to the spectrum of 29\,Psc. The \blyr\ spectra match well to the one of 29\,Psc, at all phases, red-ward of 4000\,\AA\ with the exception of the strong emission lines present in the spectrum of $\beta$\,Lyr and the significant spectral variations with respect to 29\,Psc in the UV (blue-ward of 3000 \AA).

Figure~\ref{fig:29psc_fit_phases} shows that around the eclipses the UV spectra of $\beta$\,Lyr are reasonably matched by the spectrum of 29\,Psc, while at both quadratures $\beta$\,Lyr shows a strong UV excess particularly evident around 3000-4000\,\AA. This UV excess and its phase variations is consistent with the one observed in photometry and is likely mapping the accretion stream and/or the possible accretion hot spot on the disc, both of which are only visible out of eclipse.

%\subsubsection*{Balmer \& Neutral Helium Emission Lines}

A comparison with the spectrum of 29\,Psc shows that the H$\alpha$ (6563\,\AA), H$\beta$ (4861\,\AA), He\,\textsc{i} 5876\,\AA\ and He\,\textsc{i} 6678\,\AA\ lines are strongly in emission. We have determined the equivalent width the lines, in each individual spectrum, by fitting a Gaussian of fixed FWHM corresponding to the instrumental resolution centered on the line's rest wavelength.

\begin{figure}[!ht]
    \centering
    \includegraphics[width=0.9\textwidth]{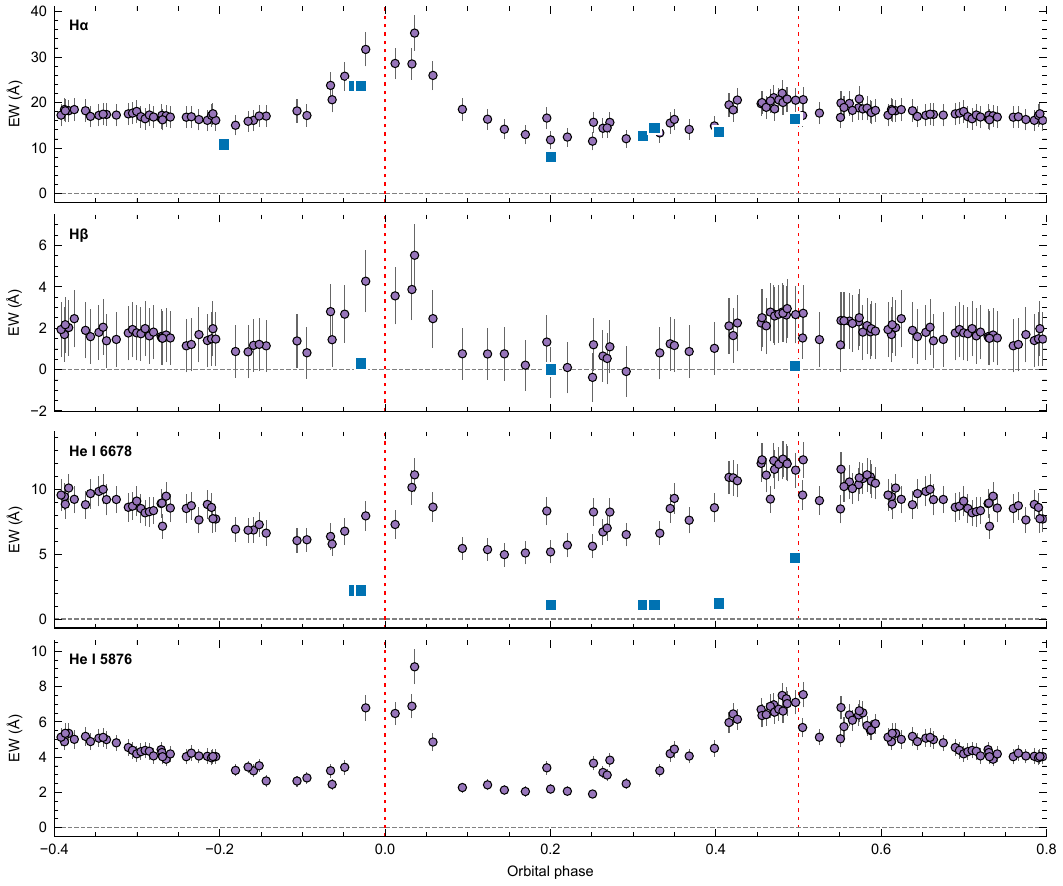}
    \caption{Measured equivalent widths of H$\alpha$, H$\beta$, He I~5876 and He I~6678 for Mauve (purple) across $\beta$ Lyrae's orbital phase, compared against the values of \citep{2011A&A...532A.148B} (blue) for H$\alpha$, H$\beta$, and He I~5876. The red dotted vertical lines mark the eclipses.}
    \label{fig:hahei6678}
\end{figure}

The phase-folded equivalent widths of H$\alpha$, H$\beta$, He\,\textsc{i}\,6678\,\AA\ and He\,\textsc{i}\,5876\,\AA\ are plotted in Fig.\,\ref{fig:hahei6678}, together with previous results for H$\alpha$ and He\,\textsc{i}\,6678\,\AA\  \cite{2011A&A...532A.148B}. Ignace et al. \citep{2018AJ....156...97I} performed a detailed, high-resolution study of the H\,$\alpha$ line, and their Fig.\,10 shows a phase-folded "line strength" plot.

The equivalent width of the emission lines varies coherently with orbital phase, exhibiting a sharper maximum at primary eclipse (phase 0.0) and a broader maximum at secondary eclipse (phase 0.5). The anti-correlation between emission-line equivalent widths (Fig.~\ref{fig:hahei6678}) and continuum brightness (Fig.~\ref{fig:phot}), together with the approximately constant H$\alpha$ line flux, supports an extended emitting region that is substantially less occulted than the stellar and disc continuum, consistent with the previously inferred bipolar outflow. As discussed by Ignace et al. \cite{2018AJ....156...97I}, who used a large number of high-resolution spectra of the H$\alpha$ line to model the emission regions in \blyr, no simple model can explain the complex variations observed (these authors modelled the emission as coming from a bipolar jet).

%\section*{Discussion} \label{sec:discussion}

The clean and highly repeatable phase plot of the emission lines, with maxima at both eclipses clearly indicate that the line-emitting gas is mostly not occulted when either component is eclipsed, placing the bulk of the emission outside the orbital plane, in agreement with the bipolar, jet-like outflow previously identified \cite{1996A&A...312..879H} and confirmed by spectropolarimetry \citep{1998AJ....115.1576H} and interferometry \citep{2008ApJ...684L..95Z, 2011A&A...532A.148B}. Bonneau et al. \citep{2011A&A...532A.148B} resolved H$\alpha$, H$\beta$ and He\,\textsc{i}~6678 emission extending at least $\sim$2\,mas ($\sim$125\,R$_\odot$) out of the orbital plane, far beyond the the stars and the thick accretion disc. They also found a compact emitting component close to the orbital plane, and a He\,\textsc{i}~6678 line-forming region that is extended at the eclipses but very small at $\phi \approx 0.3$--0.4. They attribute this to the hot spot where the gas stream from the donor meets the disc \citep{2000A&A...353.1009B}. A partly eclipsed compact component of this kind can account for the unequal heights of the two maxima. The high cycle-to-cycle reproducibility of the emission line's equivalent width indicates stability of the emitting regions over the months timeline covered by our observations, unlikely the high variability of the UV emission observed here.

The accretion disc is optically and geometrically thick and completely obscures the gainer, likely a hot star of approximately B0.5 type \cite{2018A&A...618A.112M}. The disc rim, which also contributes to the observed light, has been modelled as having a temperature of 7000--9000\,K, considerably cooler than the B6--8\,II donor ($T_{\mathrm{eff}} = 13\,300$\,K; \cite{2018A&A...618A.112M}). At primary eclipse, the disc passes in front of the hotter donor, so relatively more UV than visible light is lost, as observed. 

The deficit at secondary eclipse is harder to explain. Here, the donor passes in front of the cooler disc, which on its own would make the secondary eclipse shallower at shorter wavelengths \citep{1994ApJ...421..787K}, suggesting that something hotter than the disc rim is also hidden at this phase. A possible source is the gainer and the hot inner regions of the disc: although the disc hides the gainer, \citep{2018A&A...618A.112M} found that some light from these hot central regions still reaches us, and at secondary eclipse the donor would block it. 

The Mauve data introduce a new level of complexity in the system and in its modelling, having for the first time revealed the highly dynamical nature of some of its components. All previous work have relied on the assumption that the model is in a stable configuration, so that observational data obtained at different epochs can be combined into a single model. The UV excess emission varies roughly in phase with the equivalent widths (Fig.~\ref{fig:color_plot}), consistent with Bro\v{z} et al. \cite{2021A&A...645A..51B}. However, the cycle-to-cycle variability in the Mauve 2000-3000\,\AA\ band, first reported here, shows that the emitting region is not stable on time scales of weeks to months, possibly pointing to instabilities in the accretion flow and is indicative of a highly dynamical system, which will have to be taken into account for detailed modelling of the system. 

The results presented here show the glorious complexity of the \blyr\ system, that even though it is one of the ``classic'' variable, eclipsing stars, it is yet to relinquish all of its secrets. The uniqueness of the Mauve data resides in the broad, simultaneous wavelength coverage together with the capability to resolve and measure strong emission lines. This allows for a synoptic view of the different and numerous components of the system and to disentangle short-term variability from secular changes in the system. These components include the photospheric emission from the donor star and from the optically thick part of the disc including one or more hot bright spots, the optically thin emission from parts of the disc, from the jets and from a possible diffuse, large scale circumstellar component. The present results thus constitute a solid basis for future modelling work. At the same time, the data set presented here demonstrates how new observing facilities, even of modest size, that explore thus far unaccessed corners of the ``observational phase space'' can yield new insights, even in well-studied,  astrophysical objects that have been observed for decades.  

\section*{Methods}

\bmhead{Phase-folding}

Throughout our analysis we have used the quadratic ephemeris of \cite{2007A&A...463..233A}:

\begin{equation}
  T_{\min\mathrm{I}} = \mathrm{HJD}\;2\,408\,247.968 \;+\; 12.913\,779\,E \;+\; 3.87265\times10^{-6}\,E^{2}
  \label{eq:ephemeris}
\end{equation} 

where $T_{\min\mathrm{I}}$ refers to the time of the primary minimum and $E$ is the cycle number.

\bmhead{Frame Selection}
Mauve's observations are acquired as a sequence of ``frames'', each with an exposure of 5\,sec. To account for the still imperfect knowledge of Mauve's pointing and thus to filter out possible frames with bad pointing, for each observation, good frames are selected manually with constraints set by assuming that the total number of counts (star brightness) are not varying on the time scale of the observations (minutes, a reasonable assumption in the case of \blyr) and thus that they are distributed normally. The normality of the distribution is tested using the Pearson and Shapiro-Wilk tests. Frames that pass the test with a p-value threshold above 0.05 are selected for further processing. Over the 93 observations, 781 frames satisfied this criterion and were therefore used for our analysis, with a minimum number of three frames per observation, as required by the Shapiro-Wilk test.

\bmhead{Data Cleaning}
The raw detector counts for each frame were bias and dark current subtracted. Dark current was estimated, for each individual frame, on the basis of the count rate of the 16 pixels at either end of the detectors, that are not exposed to light and thus provide a good proxy to the detector's temperature. Each frame also carries per-pixel \texttt{QUALITY} flags produced by the standard Mauve pipeline. A pixel is considered good for a given observation if it is flagged as nominal in all selected frames. Pixels that are not flagged as nominal pixels are masked (set to \texttt{NaN}) prior to co-addition.

\bmhead{Frame Stacking}

Following on from the selection and the cleaning, the selected frames are stacked by calculating through simple averaging of the per pixel countss to produce the spectrum for a given observation. A median filter with a kernel of seven pixels is then applied, using the \texttt{medfilt} \cite{scipy_medfilt} function within the \texttt{SciPy} library.

\bmhead{Synthetic Photometry with pyphot}

Synthetic photometry of $\beta$~Lyrae was computed from the Mauve spectra using \texttt{pyphot} \cite{2026JOSS...11.8814F}, which integrates a spectrum through standard or user-defined filters. The \texttt{pyphot} software integrates the background-subtracted clean spectrum over the transmission curves of the filter. We used the standard Johnsons UBV filters, obtained through the Spanish Virtual Observatory (SVO) Filter Profile Service \cite{2020sea..confE.182R, 2026JOSS...11.8814F}, as well as two Mauve specific bands. The magnitudes for $\beta$ Lyrae are calculated by using Vega for the zero point (using an observation acquired by Mauve on the 1st of September), using:

\begin{equation}
    m_{\rm ZP} = m_{\rm Vega} + 2.5 \log_{10} \bar{F}_{\rm Vega}, 
\end{equation}

where $m_{\rm Vega}$ is the adopted magnitude of Vega in that band ($0.03$) and $\bar{F}_{\rm Vega}$ is its flux in the same filter band (obtained by integrating the Vega spectrum over the filter bandpass). The magnitude of a target then follows from its flux in the same band $\bar{F}_\star$ as

\begin{equation}
    m_\star = -2.5 \log_{10} \bar{F}_\star + m_{\rm ZP}. 
\end{equation}

The mean synthetic magnitude and colour correction terms are listed in Table~\ref{tab:colours}. The uncertainty on each magnitude is propagated from the per-pixel count errors through the filter integral. The resulting fractional statistical error is small and the error bars are smaller than the plotted symbols in Fig.\,\ref{fig:phot}. These errors are purely statistical and do not include any systematics (as, e.g., the uncertainty on the zero points, which is set by the single Vega observation and the adopted Vega magnitude), which are likely to be dominating.

In addition to the standard UBV bands, we computed magnitudes for two virtual filters in the 2000--3000 and 6000-7000\,\AA bands, using a simple top-hat shape with 100\% transmission in-band and 0\% outside. The zero point for the two non-standard bands has been derived from Vega with the same approach as the UBV bands. In this case the magnitudes depend on the band definition and are only meaningful relative to Vega, and so cannot be compared directly with other photometric systems. However, the zero point is the same for all observations, so the changes in magnitude over the orbit, such as the eclipse depths, are independent of it. 

For comparison with the historical UV observations performed by the Orbiting Astronomical Observatory (OAO) A-2 and IUE satellites \citep{1994ApJ...421..787K}, we computed the primary and secondary eclipse depth, in magnitudes, for the Mauve 2000-3000\,\AA\ band. For each of primary minimum, secondary minimum and maximum (phases 0.25 and 0.75), we took the median magnitude of the observation within $\pm0.05$ in phase, and defined each eclipse depth as the difference between the minimum and the maximum. For the historical observations, the depths were estimated from the published data tables of \citep{1994ApJ...421..787K}, which list magnitudes relative to maximum light, as the faintest value near each eclipse. The results (Table~\ref{tab:mauve_unique}) show similar eclipse depths in the Mauve 2000-3000\,\AA\ band and in the narrower historical UV observation bands \citep{1994ApJ...421..787K}. Both primary and secondary eclipses are slightly deeper in the UV than in the red.

\bmhead{Comparison to 29\,Psc}

To measure the emission lines and assess the presence of a continuum excess in \blyr\ with respect to the photospheric spectrum, we have compared it with the spectrum of 29\,Psc observed with the same instrument. Eight observations of 29\,Psc, obtained between 28 August and 23 September, were each reduced as for \blyr\ using the methods aforelisted, and averaged. Because 29\,Psc and \blyr\ differ somewhat in colour, the 29\,Psc spectrum has been scaled to each \blyr\ spectrum by a factor $a + b\lambda$, where $a$ and $b$ were determined through a least squares process with iterative 2.5$\sigma$ clipping over 3800--7000\,\AA, excluding $\pm$80--120\,\AA\ around H$\gamma$, H$\beta$, He\,\textsc{i}~5876, H$\alpha$ and He\,\textsc{i}~6678. This process effectively matches the slope of the continuum for 29\,Psc\ to the one of \blyr, effectively compensating for the effective temperature difference. We refer to this scaled spectrum as the fitted 29\,Psc spectrum. Both spectra are median-filtered over 9 pixels before the fit. The scaled spectrum matches very well the photospheric continuum of \blyr, so that the difference or ratio between \blyr\ and the fitted 29\,Psc shows the presence the line emission and any continuum excess or deficit. Below 3800\,\AA\ the fit is extrapolated, so the UV excess is measured relative to the optical continuum. We have compared the fitted 29\,Psc\ spectrum to the phase-averaged \blyr\ spectrum (the mean of 20 equal phase bins, each weighted equally), to the mean spectra at selected phases, and to each individual spectrum for the equivalent-width measurements. Fig.\,\ref{fig:29psc_fit} shows the resulting fit and the difference.

\bmhead{Determination of the emission lines' equivalent width}

The equivalent widths (EW) of H$\alpha$, H$\beta$, He\,\textsc{i}~5876 and He\,\textsc{i}~6678 are measured for each spectrum by fitting the residual
\begin{equation}
  r(\lambda) = \frac{F(\lambda) - M(\lambda)}{C(\lambda)}
             = a + b\,(\lambda - \bar\lambda) + \sum_i \mathrm{EW}_i\, G(\lambda - \lambda_i),
  \qquad
  G(x) = \frac{e^{-x^2/2\sigma^2}}{\sigma\sqrt{2\pi}},
  \label{eq:ew}
\end{equation}
where $F(\lambda)$ is the observed $\beta$~Lyrae spectrum, $M(\lambda)$ is the mean 29~Psc spectrum scaled by a first-order polynomial fitted to $F(\lambda)$, and $C(\lambda)$ is a straight-line continuum through windows on either side of each line (Table~\ref{tab:linewindow}). Each line is a unit-area Gaussian $G$ with the instrumental FWHM of 75\,\AA\ ($\sigma = \mathrm{FWHM}/2\sqrt{2\ln 2}$), centred at the rest wavelength $\lambda_i$, on a linear baseline $a + b(\lambda - \bar\lambda)$. Table \,\ref{tab:linewindow} shows the rest wavelength and line window values, with the fit displayed in Fig.\,\ref{fig:ew_check}. The fitted amplitude $\mathrm{EW}_i$ is therefore the equivalent width of the emission above the absorption lines of 29\,Psc. H$\alpha$ and He\,\textsc{i}~6678 are fitted together, as their profiles overlap. Subtracting $M(\lambda)$ reintroduces the photospheric absorption of 29~Psc into the residual. The net equivalent width is therefore

\begin{equation}
  \mathrm{EW}_{\mathrm{net},i} = \mathrm{EW}_i - \left|\mathrm{EW}_{\mathrm{abs},i}\right|,
  \label{eq:ewnet}
\end{equation}

where $\mathrm{EW}_{\mathrm{abs},i}$ is measured by fitting Eq.~\ref{eq:ew} to a synthetic 29\,Psc spectrum obtained from the CALSPEC database \cite{2020AJ....160...21B} and smoothed to the Mauve resolution. No correction is applied to He\,\textsc{i}~5876, where the 29\,Psc spectrum shows no measurable absorption. Uncertainties combine the fit error with a systematic term from varying the FWHM (65--85\,\AA) and line centres ($\pm$5\,\AA), and the uncertainty of the absorption correction. 

The EWs are calculated in the phase bins around the primary eclipse (0.9--1.1), after primary eclipse (0.1--0.4), around secondary eclipse (0.4--0.6), and after secondary eclipse (0.6--0.9). We compared the results to the values obtained by \cite{1998AJ....115.1576H} in Table \,\ref{tab:hoffman}. A simple geometrical model of the system based on  \cite{2013MNRAS.432..799M}, is shown included in Fig.\,\ref{fig:geometry} to help visualise how the line emission is modulated.

The equivalent width concept relies on the measurement of the intensity of the emission against the a continuum. If the two are however originating from different regions (and thus physically not related) the measurement may not be physically meaningful. To test whether the equivalent-width variation reflects changes in the eclipsed continuum rather than in the line emission, we compared the mean Mauve spectra (within $\pm0.05$ in phase) at primary minimum ($\phi=0$) and at first quadrature ($\phi=0.25$). Each spectrum was normalised to the 29\,Psc spectrum with the same process as above, and then both spectra were normalised to the continuum at primary minimum (Fig.\,\ref{fig:linevscontinuum}). This procedure shows that the continuum increases by a factor of $2.16\pm0.09$, whereas the H$\alpha$ line flux increases by only $1.20\pm0.06$, so that the variation of the flux in the H$\alpha$ emission is small, likely pointing to the emission coming from a large, diffuse region little affected by the eclipses.

\section*{Supplementary Material}
\begin{figure}[!ht]
    \centering
    \includegraphics[width=0.9\textwidth]{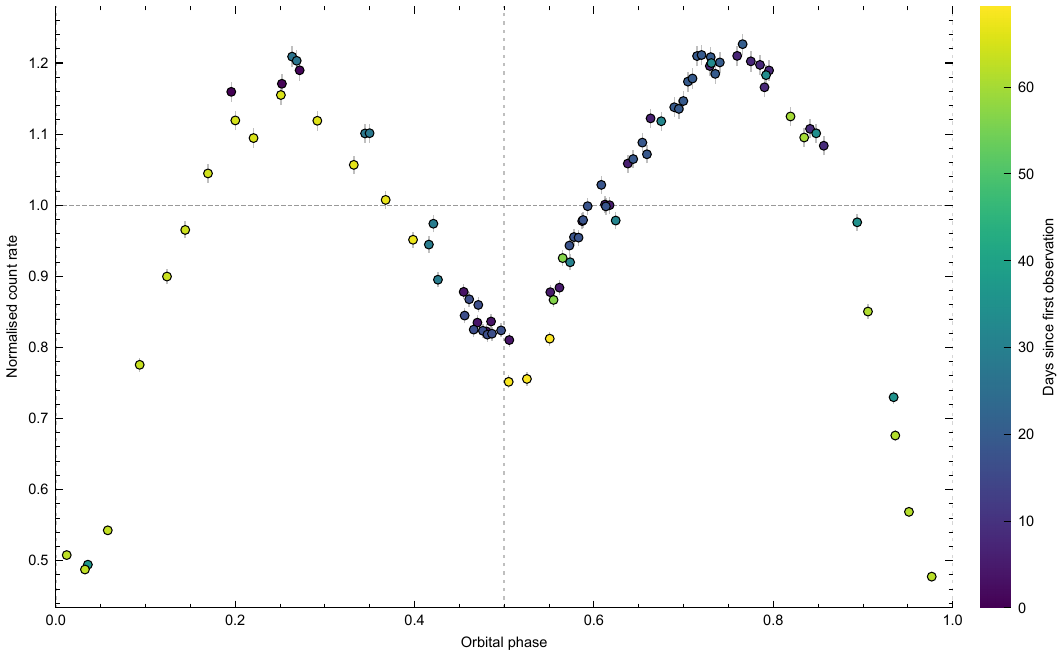}
    \caption{Normalised count rate (counts/sec) against orbital phase of $\beta$ Lyrae, coloured by days since first observation. The data collection campaign ran from the 5th of August to the 4th of October.}
    \label{fig:lightcurve}
\end{figure}

\begin{figure}[!ht]
    \centering
    \includegraphics[width=0.9\textwidth]{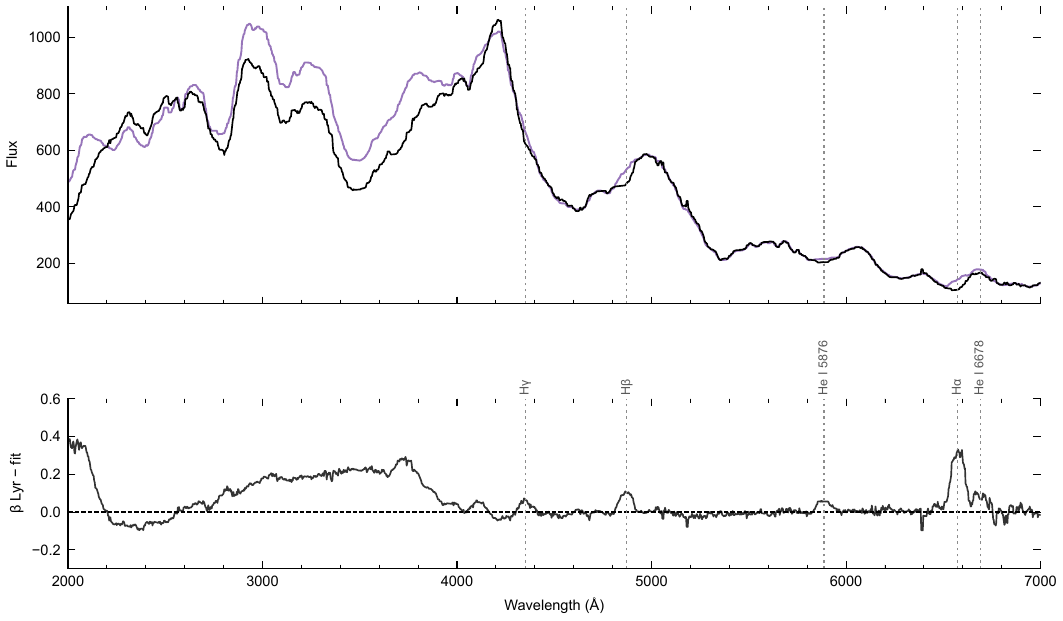}
    \caption{Mean Mauve spectrum of $\beta$~Lyrae (purple; phase-balanced mean of 20 phase bins) compared with a fitted mean spectrum of the B7 comparison star 29~Psc (black). Top: observed and fitted spectra. Bottom: difference between the two. Dashed lines mark H$\alpha$, H$\beta$, H$\gamma$, He\,I~5876\,\AA\,\ and He\,I~6678\,\AA\,, after applying a $+1.15$\,nm shift to the wavelength scale so that the lines fall at their expected wavelengths.}
    \label{fig:29psc_fit}
\end{figure}

\begin{figure}[!ht]
    \centering
    \includegraphics[width=1.0\textwidth]{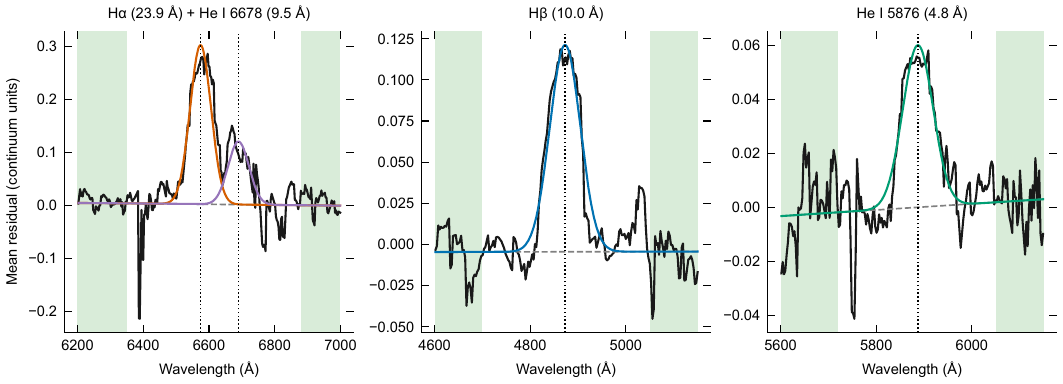}
    \caption{Mean residual of the $\beta$~Lyrae spectra after subtracting the fitted 29~Psc spectrum, in continuum units (black), with the fitted Gaussian line profiles (coloured; FWHM 75\,\AA, fixed centres, dotted) on a linear baseline (dashed). The EW of each line is given above its panel. Green bands mark the continuum windows.}
    \label{fig:ew_check}
\end{figure}

\begin{figure}[!ht]
    \centering
    \includegraphics[width=1.0\textwidth]{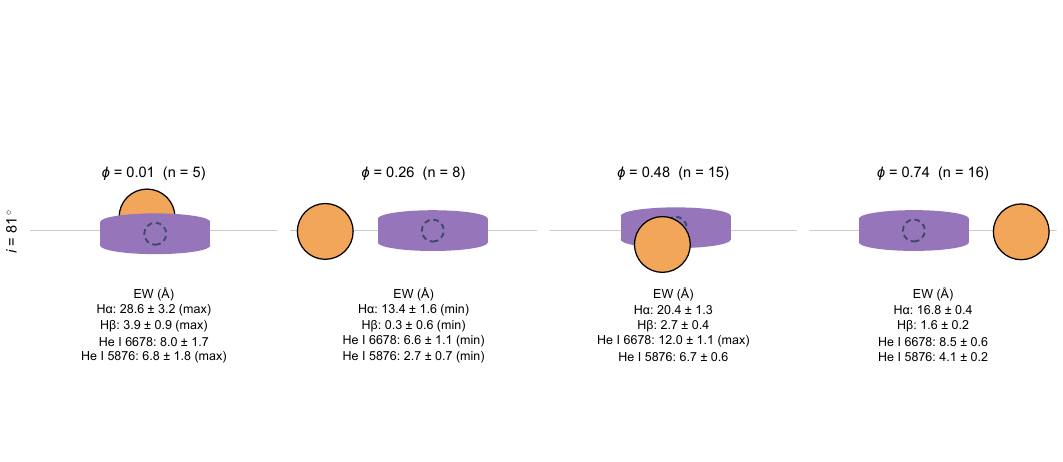}
    \caption{Schematic representation of the geometry of the $\beta$\,Lyrae system at the four phases near 0, 0.25, 0.5, and 0.75, using inclination ($i=81$) based on the model of \citep{2013MNRAS.432..799M}. n represents the number of spectra included in the phase bin within $\pm 0.05$.}
    \label{fig:geometry}
\end{figure}

\begin{figure}[!ht]
    \centering
    \includegraphics[width=0.8\textwidth]{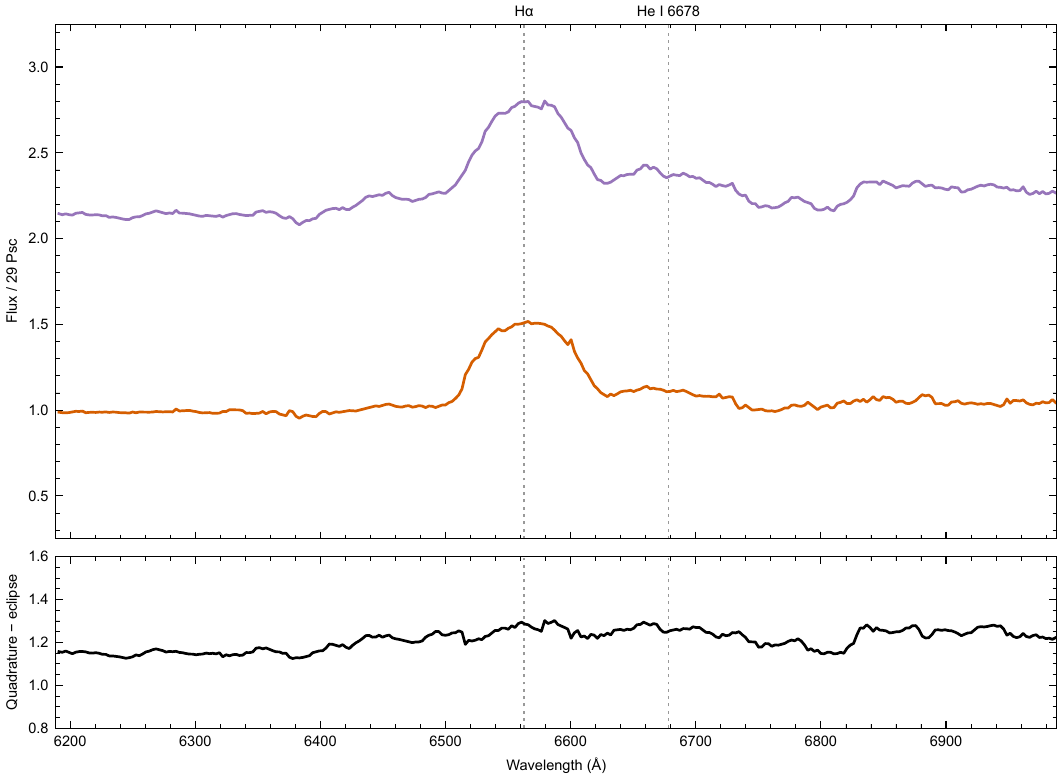}
    \caption{Mauve spectra of $\beta$~Lyrae around H$\alpha$ at quadrature (purple; $\approx0.25$) and primary eclipse (orange; $\approx0$), divided by the 29~Psc spectrum (top), and their difference (bottom). The continuum changes by a factor of $\approx$2.2, while the H$\alpha$ flux stays nearly constant.}
    \label{fig:linevscontinuum}
\end{figure}

\bmhead{Supplementary Tables and Figures}
Table~\ref{tab:colours},~\ref{tab:mauve_unique},~\ref{tab:windows},~\ref{tab:hoffman} and Fig.~\ref{fig:lightcurve},~\ref{fig:29psc_fit},~\ref{fig:ew_check},~\ref{fig:geometry},~\ref{fig:linevscontinuum}. 

\begin{table}[!ht]
\caption{Mean synthetic magnitudes and colours of $\beta$~Lyrae in the Johnson system (error: standard error of the mean).}\label{tab:colours}
\begin{tabular}{@{}llll@{}}
\toprule%
Colour & Magnitude \\
\midrule
U & $3.036 \pm 0.026$ \\
B & $3.545 \pm 0.024$ \\
V & $3.609 \pm 0.023$ \\
\midrule
U$-$B & $-0.509 \pm 0.003$ \\
B$-$V & $-0.064 \pm 0.002$ \\
\botrule
\end{tabular}
\end{table}

\begin{table}[!ht]
\caption{Synthetic photometry of $\beta$~Lyrae in the two Mauve-unique
bands, compared with mid-UV eclipse depths from OAO\,A-2 and IUE
\citep{1994ApJ...421..787K}. Mauve magnitudes are relative to Vega; the
mean is taken over all observations, and eclipse depths are measured
relative to maximum light. Literature depths are the faintest relative
magnitudes near each eclipse and are approximate.}
\label{tab:mauve_unique}
\begin{tabular*}{\textwidth}{@{\extracolsep\fill}llcccc}
\toprule
Source & Band & Mean & Maximum & Primary depth & Secondary depth \\
       &      & (mag) & (mag)  & (mag)         & (mag)           \\
\midrule
Mauve & 2000--3000\,\AA\, & 2.83 & 2.55 & 0.96 & 0.46 \\
Mauve & 6000--7000\,\AA\, & 3.53 & 3.32 & 0.87 & 0.39 \\
\midrule
IUE    & 2180\,\AA\, & -- & -- & $\sim$0.84 & $\sim$0.4  \\
OAO~A-2 & 2460\,\AA\, & -- & -- & $\sim$0.88 & $\sim$0.47 \\
IUE    & 2470\,\AA\, & -- & -- & $\sim$0.99 & $\sim$0.5  \\
OAO~A-2 & 2980\,\AA\, & -- & -- & $\sim$1.03 & $\sim$0.49 \\
\botrule
\end{tabular*}
\end{table}

\begin{table}
\caption{Line windows used for the equivalent width measurements.}\label{tab:windows}
\begin{tabular*}{\textwidth}{@{\extracolsep\fill}lcc}
\toprule%
Line & Rest wavelength (\AA\,) & Line window (\AA\,) \\
\midrule
H$\alpha$ & 6562.8 & 6513.3--6635.3 \\
H$\beta$ & 4861.3 & 4811.8--4933.8 \\
He\,\textsc{i}~6678 & 6678.2 & 6664.7--6714.7 \\
He\,\textsc{i}~5876 & 5875.6 & 5827.1--5947.1 \\
\botrule
\label{tab:linewindow}
\end{tabular*}
\end{table}

\begin{table}[!ht]
\caption{Mauve equivalent widths in the phase bins of Hoffman et al.\ (1998) \cite{1998AJ....115.1576H}, compared with their values (corrected for underlying absorption for H$\alpha$ and H$\beta$). Mauve values are the mean $\pm$ standard error of the individual spectra in each bin. PE: primary eclipse (0.9--1.1); APE: after primary eclipse (0.1--0.4); SE: secondary eclipse (0.4--0.6); ASE: after secondary eclipse (0.6--0.9).}\label{tab:hoffman}
\begin{tabular*}{\textwidth}{@{\extracolsep\fill}lccc}
\toprule%
Line, bin & Mauve (\AA) & Hoffman et al.\ (\AA) & Mauve / Hoffman \\
\midrule
H$\alpha$, PE & $25.6 \pm 3.6$ & 28.6 & $0.89 \pm 0.13$ \\
H$\alpha$, APE & $14.2 \pm 2.2$ & 12.4 & $1.15 \pm 0.18$ \\
H$\alpha$, SE & $19.4 \pm 2.5$ & 15.1 & $1.28 \pm 0.17$ \\
H$\alpha$, ASE & $17.1 \pm 2.4$ & 12.9 & $1.33 \pm 0.18$ \\
\midrule
H$\beta$, PE & $2.8 \pm 1.4$ & 3.9 & $0.72 \pm 0.37$ \\
H$\beta$, APE & $0.7 \pm 1.2$ & 1.0 & $0.66 \pm 1.24$ \\
H$\beta$, SE & $2.3 \pm 1.3$ & 1.3 & $1.74 \pm 1.03$ \\
H$\beta$, ASE & $1.6 \pm 1.3$ & 0.8 & $2.02 \pm 1.60$ \\
\midrule
He\,\textsc{i}~6678, PE & $7.6 \pm 1.0$ & 4.9 & $1.55 \pm 0.21$ \\
He\,\textsc{i}~6678, APE & $6.9 \pm 0.8$ & 2.3 & $3.01 \pm 0.35$ \\
He\,\textsc{i}~6678, SE & $11.0 \pm 1.1$ & 4.5 & $2.45 \pm 0.25$ \\
He\,\textsc{i}~6678, ASE & $8.5 \pm 0.9$ & 2.9 & $2.93 \pm 0.30$ \\
\midrule
He\,\textsc{i}~5876, PE & $4.8 \pm 0.8$ & 9.0 & $0.54 \pm 0.09$ \\
He\,\textsc{i}~5876, APE & $3.1 \pm 0.4$ & 3.6 & $0.86 \pm 0.10$ \\
He\,\textsc{i}~5876, SE & $6.4 \pm 0.6$ & 6.4 & $1.00 \pm 0.09$ \\
He\,\textsc{i}~5876, ASE & $4.3 \pm 0.4$ & 3.3 & $1.29 \pm 0.12$ \\
\botrule
\end{tabular*}
\end{table}

%If your article has accompanying supplementary file/s please state so here. 

%Authors reporting data from electrophoretic gels and blots should supply the full unprocessed scans for key as part of their Supplementary information. This may be requested by the editorial team/s if it is missing.

%Please refer to Journal-level guidance for any specific requirements.

\backmatter

\bmhead{Acknowledgements}
We thank C3S LLC, ISISPACE BV, MediaLario S.r.l., Avantes BV, and CeramOptec GmbH, for their partnership with Blue Skies Space in constructing and delivering Mauve.

Views and opinions expressed here those of the author(s) only and do not necessarily reflect those of the European Union or the European Research Council. Neither the European Union nor the granting authority can be held responsible for them.

This research acknowledges the use of the Spanish Virtual Observatory (SVO) Filter Profile Service (\url{https://svo2.cab.inta-csic.es/theory/fps/index.php}) and the CALSPEC database (\url{https://www.stsci.edu/hst/instrumentation/reference-data-for-calibration-and-tools/astronomical-catalogs/calspec}).

\section*{Declarations}

\bmhead{Funding}

The Mauve project has received funding from the European Union’s Horizon Europe research and innovation programme under grant agreement No. 101082738, and was supported by the UK Research and Innovation (UKRI)'s Horizon Europe Guarantee Scheme. Krisztián Vida, Bálint Seli, and Blanka Schmercz's research was funded by the Hungarian National Research, Development and Innovation Office \'Elvonal grant KKP-143986. Emma T. Whelan, Patrick F. Flanagan’s, and Jamie J. Stewart's memberships of the Mauve Science Collaboration is supported by Taighde ÉIREANN - Research Ireland under Grant number 24/FFP-P/12957. Elena Pancino and Sharmila Rani are co-funded by the European Union (ERC-2022-AdG, {\em "StarDance: the non-canonical evolution of stars in clusters"}, Grant Agreement 101093572, PI: E. Pancino).

\bmhead{Competing interests}
The authors declare no competing interests.

\bmhead{Ethics approval and consent to participate}
Not applicable.

\bmhead{Consent for publication}
Not applicable.

\bmhead{Data availability}
The Mauve spectra of $\beta$ Lyrae, 29 Psc, and Vega used in this work will be made publicly available on Blue Skies Space’s Stardrive platform and the Barbara A. Mikulski Archive for Space Telescopes (MAST) archive.

\bmhead{Materials availability}
Not applicable.

\bmhead{Code availability}
This work made use of NumPy \citep{Harris2020}, SciPy \citep{Virtanen2020}, Astropy \citep{Astropy2013, Astropy2018, Astropy2022}, Matplotlib \citep{Hunter2007} and pyphot \citep{2026JOSS...11.8814F}.

\bmhead{Author contribution}
S.Raz, I.G., F.F. led the analysis of the data and the astrophysical interpretation. B.J.W., S.Raz, A.S., H.J., I.S., P.W. contributed to the scheduling, acquisition and verification of the observations presented here. J.T., A.S., M.T., G.T., B.J.W., L.J.B., Y.B.R.M. contributed to the preparation of the manuscript. All other authors provided notable feedback.

% \noindent
% If any of the sections are not relevant to your manuscript, please include the heading and write `Not applicable' for that section. 





\bibliography{sn-bibliography}% common bib file

\end{document}